\documentstyle[12pt]{article}
\newcommand{\newc}{\newcommand}
\newc{\beq}{\begin{equation}}
\newc{\eeq}{\end{equation}}
\newc{\bea}{\begin{eqnarray}}
\newc{\eea}{\end{eqnarray}}
\newc{\mpl}{M_P}
\newc{\mgrav}{m_{3/2}}
\newc{\mgut}{M_{\rm G}}
\newc{\mstring}{M_{\rm string}}
\newc{\mw}{m_W}
\newc{\msusy}{M_{\rm SUSY}}
\newc{\mint}{M_{\rm int}}
\newc{\gev}{\,\mbox{GeV}}
\newc{\tr}{\mbox{Tr}\,}
\renewcommand{\bar}{\overline}
\newc{\Si}{\Sigma}
\newc{\eps}{\epsilon}
\newc{\ie}{{\it i.e.\/}}
\newc{\eg}{{\it e.g.\/}}
\newc{\gsim}{\lower.7ex\hbox{$\;\stackrel{\textstyle>}{\sim}\;$}}
\newc{\lsim}{\lower.7ex\hbox{$\;\stackrel{\textstyle<}{\sim}\;$}}
\def\NPB#1#2#3{{\it Nucl.\ Phys.} {\bf B#1} (19#2) #3}
\def\PLB#1#2#3{{\it Phys.\ Lett.} {\bf B#1} (19#2) #3}
\def\PLBold#1#2#3{{\it Phys.\ Lett.} {\bf#1B} (19#2) #3}
\def\PRD#1#2#3{{\it Phys.\ Rev.} {\bf D#1} (19#2) #3}
\def\PRL#1#2#3{{\it Phys.\ Rev.\ Lett.} {\bf#1} (19#2) #3}
\def\PRT#1#2#3{{\it Phys.\ Rep.} {\bf#1} (19#2) #3}

\begin{document}

\begin{titlepage}
\begin{flushright}
{\rm
IASSNS-HEP-97-137\\
CERN-TH/98-75\\
RU-97-97\\
{\tt hep-ph/9803310}\\
March 1998\\
}
\end{flushright}
\vskip 2cm
\begin{center}
{\Large\bf Stabilized Singlets in Supergravity as a Source of the
$\mu$-parameter}
\vskip 1cm
{\large
Christopher Kolda ${}^1$, Stefan Pokorski ${}^{2,3}$ and Nir Polonsky ${}^4$\\}
\vskip 0.4cm
{\small\it ${}^1$ School of Natural Sciences, Institute for Advanced Study,
Princeton, NJ 08540, USA\\
${}^2$ CERN Theory Division, CH-1211 Geneva 23, Switzerland \\
${}^3$ Institute of Theoretical Physics, Warsaw University, H\.oza 69,
00-681 Warsaw, Poland\\
${}^4$ Department of Physics and Astronomy, Rutgers University, Piscataway,
NJ 08854, USA\\
}
\end{center}
\vskip .5cm
\begin{abstract}
Within the context of supergravity-coupled supersymmetry, fields which
are gauge and global singlets are usually considered
anathema. Their vacuum expectation values are shifted by quadratically 
divergent tadpole diagrams which are cutoff at the Planck scale, destabilizing
the classical potential and driving the singlet field to large values.
We demonstrate a new and generic mechanism which stabilizes
the singlet in the presence of an extended gauge symmetry. Such a
symmetry will be broken down to the Standard Model
by the supergravity interactions
near the scale of spontaneous supersymmetry-breaking in the
hidden-sector ($\sim10^{10-11}\gev$). The resulting singlet
expectation value is stabilized and naturally of order the gravitino mass,
providing therefore a weak-scale mass for the Higgs fields
of the supersymmetric Standard Model (a $\mu$-parameter). The
resulting low-energy theory is the minimal supersymmetric Standard
Model, with all new fields decoupling at the intermediate scale.
\end{abstract}
\end{titlepage}
\setcounter{footnote}{0}
\setcounter{page}{1}
\setcounter{section}{0}
\setcounter{subsection}{0}
\setcounter{subsubsection}{0}


One of the primary motivations for low-energy supersymmetry (SUSY) is
the stabilization of the
weak scale ($\mw$) against quadratic divergences which
appear at the quantum level and can drive $\mw$ up to the
Planck scale ($\mpl$). 
To accomplish such a feat, one typically postulates that there are two
separate sectors in the Lagrangian describing all of physics: a visible sector
in which exist the Standard Model (SM) and its SUSY
extension, the Minimal Supersymmetric Standard Model (MSSM); and a hidden
sector in which exist the fields and interactions necessary to spontaneously
break SUSY (see \eg~\cite{nilles}). Consistency of the model requires that
the hidden and visible sectors must not communicate with one
another directly via renormalizable, tree-level interactions, and so
one must also postulate a mediation mechanism which communicates SUSY-breaking
from the hidden to the visible sector.

The simplest and most elegant means for communicating SUSY-breaking is via
supergravity interactions; in fact, this mechanism is
inherited directly from the gauging of SUSY and requires no additional
sectors or interactions. In the picture of supergravity mediation, SUSY is
spontaneously
broken in the hidden sector at a scale $\msusy$.
This scale is then communicated to the visible sector through Planck-scale
suppressed operators so that the visible SUSY-breaking scale is 
$\msusy^2/\mpl$. Setting this scale to $\mw$ (in order to have weak-scale
SUSY), one finds that the hidden
sector scale is then $\msusy\simeq\sqrt{\mw\mpl}\equiv\mint\simeq10^{10-11}
\gev$. The
resulting picture has three scales (though they are not all independent): 
$\mpl$, $\mint$ and $\mw$. The physics in each sector, however, is
only sensitive to two of these scales: $\mpl$ and $\mint$ in the
hidden sector, and $\mpl$ and $\mw$ in the visible sector. 
In both cases, one expects that SUSY-preserving masses should be near
$\mpl$, and that SUSY-breaking masses in each sector should be near the scale
which controls SUSY-breaking in that sector.

However, there is one necessary, and another possible, violation of these
expectations in low-energy SUSY.
The former is often referred to as the $\mu$-problem~\cite{muproblem}. 
The superpotential of the MSSM must for
phenomenological reasons contain a term $\mu H\bar H$ which mixes the two
Higgs
doublets responsible for electroweak symmetry breaking, with $\mu\sim\mw$.
However, $\mu$ is both SUSY- and $SU(2)\times U(1)$-invariant and so we 
expect that its natural scale should be the Planck scale. The other
possible violation
arises whenever one wishes to extend the gauge structure of the MSSM to include
new interactions broken at some scale intermediate to $\mw$ and $\mpl$. In
particular, there exist several arguments that there should be new
physics in the visible sector at a scale close to 
$\mint\sim10^{10-11}\gev$ (for example, 
a right-handed neutrino at such a scale could provide a suppressed
neutrino mass through the
seesaw mechanism consistent with current indications).
It is remarkable that this new scale for visible sector physics 
is also roughly the scale of SUSY-breaking in the hidden
sector, though there is no obvious connection. In fact we will show in
this letter that there
may indeed be a deep underlying connection. Specifically, we will
demonstrate a mechanism which simultaneously solves the $\mu$-problem
while generating a scale for new visible sector physics close to the
scale $\msusy$ of hidden sector physics.

Of the existing suggestions for solving the $\mu$-problem of the MSSM,
two bear some kinship to our proposal. The first such suggestion
is to extend the matter content of the MSSM by a singlet, $S$, which
couples to $H\bar H$. If $S$ receives a weak-scale vacuum expectation value
(vev), it dynamically provides a $\mu$-parameter, $\mu\sim\left\langle
S\right\rangle$. This is known as the
Next-to-Minimal SUSY Standard Model (NMSSM)~\cite{nmssm}. The second solution
is the Giudice-Masiero mechanism~\cite{gm} in which non-minimal,
non-renormalizable K\"ahler
couplings mix the hidden and visible sectors. Thus what appears to be a
SUSY-preserving $\mu$-parameter originates in fact from SUSY-breaking in
the hidden sector. Our
mechanism has ingredients from both proposals in that we add a gauge singlet to
the spectrum along with a non-minimal K\"ahler potential. But the
singlet will decouple from the low-energy theory, and no potentially dangerous
mixing of hidden
and visible sectors is necessary. Our proposal is more closely related to that
of Sen (and as updated by Barr)~\cite{sen} for reviving the ``sliding singlet''
solution for the doublet-triplet splitting problem of $SU(5)$. There, as here,
dangerous destabilizing operators in the effective potential need to be
suppressed; in Ref.~\cite{sen}, 
this is accomplished by extending $SU(5)$ in a very
particular way. Their models appear, in fact, to be special cases of a more
general principle which we will identify and explore here.

Begin by considering the NMSSM, which has two well-known problems.
At the renormalizable level, the NMSSM has a $Z_3$
symmetry. If that symmetry is preserved to all orders, then the vev of $S$ will
break the symmetry at the weak scale and produce cosmologically dangerous
domain walls. If on the other hand the $Z_3$ symmetry is not preserved by
higher-order terms in the Lagrangian, then $S$ carries no conserved quantum
numbers. In this latter case, $S$ will generically
develop tadpoles, in the presence of spontaneously-broken
SUSY, whose quadratic divergences are cut
off by the Planck scale~\cite{old,bp}. The resulting 
shift in the potential for $S$ causes it to slide to large values far
above the weak scale. If it were to couple to the MSSM Higgs fields,
they would receive unacceptably large masses, destabilizing the weak
scale. Therefore one concludes that not only do singlets fail to
provide a viable $\mu$-parameter, but they cannot even be allowed to
couple to light fields.

We now present a toy model which solves this destabilization problem of the
NMSSM, while at the same time introducing a new
visible sector interaction whose scale will naturally fall at $\mint$.
Within this model, singlets can couple to MSSM fields, and in
particular, can provide dynamical $\mu$-terms at the weak scale.
The model itself
demonstrates a very general mechanism, though it already contains all of the
ingredients necessary to be phenomenologically viable. 

~

\noindent {\it The Model.}~~ Consider a superpotential
\beq
W=\lambda_H SH\bar H + \lambda_\Sigma S\Si\bar\Si
\eeq
where $H,\bar H$ carry charges $\pm1$ under a gauge symmetry $U(1)_H$, 
$\Si,\bar\Si$ are charged $\pm1$ under another gauge
symmetry $U(1)_\Si$, and $S$ is a gauge singlet. We require $\Si,\bar\Si$ to
be neutral under $U(1)_H$ and, for simplicity, assume that $H,\bar H$
are also neutral under $U(1)_\Si$, though they need not be. The
$D$-term for $U(1)_\Si$ is then simply
\beq
D_\Si=g_\Si\left(|\Si|^2-|\bar\Si|^2\right),
\eeq
and similarly for $D_H$. To apply this toy model to the MSSM, we identify
$H,\bar H$ as the usual Higgs doublets, and extend $U(1)_H$ to the Standard Model
gauge group; $\Si$ and $\bar\Si$ are new fields charged under a new gauge
symmetry $U(1)_\Si$.

At the level of the superpotential, there exists the usual $Z_3$ which
forbids explicit mass terms from appearing in $W$~\cite{nmssm}. 
This symmetry is broken
by $S\neq0$, which could lead to creation of 
electroweak scale domain walls via the Kibble mechanism. 
The appearance of an $S^3$ term is
forbidden by an $R$-symmetry under which $R(W)=2$ and $R(S)=0$.

However, we will assume that the $Z_3$ symmetry of the superpotential is
only an accidental symmetry. This is a natural
expectation since global symmetries are generally not 
preserved by quantum gravity effects (unless they are remnants of 
broken gauge symmetries)~\cite{global}. 
In particular, we expect that gravity-induced global symmetry-breaking will
appear as non-renormalizable, explicit symmetry-breaking terms in the K\"ahler
potential, $K$. 
The $Z_3$ symmetry can thus be a symmetry of the effective
superpotential without being a symmetry of the entire action.
This is equivalent to the statement that the $S$ field is a true 
singlet, carrying no conserved quantum numbers. 

After SUSY-breaking, non-zero tadpoles for $S$ will generically arise with 
light chiral fields circulating in the loops~\cite{old,bp,jain,bpr,np}. 
These tadpoles appear
at ${\cal O}(\hbar^n/\mpl)$ due to supergravity corrections from
Planck-suppressed operators. Because the exact source of the couplings which
generate the tadpoles is highly model-dependent, we do not know {\it a priori}
at what loop order non-zero contributions are generated. For example, it is
known that for a flat K\"ahler metric, non-zero tadpoles do not arise
until two-loops~\cite{jain,bpr}; however, for a non-flat metric they may 
arise at one-loop.

We will analyze these tadpole contributions following the formalism laid out in
Refs.~\cite{bp,bpr}. Putting aside the question of loop-order at which the
tadpoles arise, it is sufficient to consider these contributions as coming from
an
effective K\"ahler potential of the form
\beq
K=\left\{1+\frac{1}{\mpl}(S+S^\dagger)\right\}\Phi_i\Phi_i^\dagger.
\eeq
The first term is just the canonical contribution to $K$ of a chiral 
superfield $\Phi_i$, 
while the second term explicitly breaks the $Z_3$-symmetry
present in the superpotential.
In general, an $n$-loop induced tadpole generates a term in the effective
Lagrangian of the form:
\beq
{\cal L}_{{\rm eff}}
\sim \frac{N}{(16\pi^2)^n}\frac{\Lambda^2}{\mpl}\int d^4\theta\,
e^{K/\mpl^2}\,(S+S^\dagger)
\eeq
where $N$ counts the number of light chiral superfields that appear in the
loops and $\Lambda$ is the cutoff for the quadratic divergence.
(This expression results from a full supergravity calculation~\cite{bpr}.)
Henceforth we make the reasonable assumption that $\Lambda\simeq\mpl$.

After SUSY-breaking, the superspace density, $e^{K/\mpl^2}$, has the expansion:
\beq
e^{K/\mpl^2}=1+\frac{1}{\mpl^2}\left\{\theta^2 K_iF^i+{\bar\theta}^2 
K_{\bar\imath}
F^{\bar\imath} +\theta^2{\bar\theta}^2\left(K_{i\bar\jmath}+\frac{K_i 
K_{\bar\jmath}}
{\mpl^2}\right) F^i F^{\bar\jmath}\right\}
\label{density}
\eeq
where $K_i$ $(K_{\bar\imath})$ is the derivative of the K\"ahler potential
with respect to a superfield $\Phi_i$ $(\Phi^\dagger_i)$. Even though $i,j$ are
summed over
all fields in the theory, the vev of the superspace density is obviously
dominated by the fields responsible for SUSY-breaking. In particular, we can
take $F^i\sim \msusy^2$ which is the scale of SUSY-breaking in the
hidden sector and is related to the gravitino mass, $\mgrav$, via
$\msusy^2\simeq\mgrav\mpl$ (assuming cancellation of the cosmological
constant).
Throughout this article we will always assume that the
communication of SUSY-breaking to the visible sector is via
supergravity-induced terms, and thus $\mgrav\simeq\mw$. (See Ref.~\cite{np}\ for
the case of tadpoles in gauge-mediated models for which $\mgrav\ll\mw$.)

While $K_{ij}\sim{\cal O}(1)$,
the value of $K_i$ is more model-dependent. For example, in the
Polonyi model (see~\cite{nilles}),
the field responsible for SUSY-breaking has an $F$-term $\sim\mw\mpl$,
but its scalar vev is $\sim\mpl$; thus $K_i\sim\mpl$. However, this is not
a necessary ingredient and $K_i\ll\mpl$ is possible. To be general we will
write
\beq
K_i=\eps_i\mpl \quad\mbox{for}\quad |\eps_i|\leq1
\eeq
to parameterize our ignorance of the value of $K_i$.
The contribution to the effective Lagrangian coming from the tadpole is then:
\bea
{\cal L}_{{\rm eff}}&\sim&\frac{N}{(16\pi^2)^n}
\left\{K_{\bar\imath}\mgrav F_S+\left(K_{i\bar\jmath}
+\frac{K_i K_{\bar\jmath}}{\mpl^2}
\right)\mgrav^2\mpl S + h.c.\right\} \nonumber \\
&\equiv&\beta\eps\mgrav\mpl F_S + \gamma\mgrav^2\mpl S + h.c.
\label{eq:leff}
\eea
where $\eps$ is the maximum $\eps_i$ among the SUSY-breaking fields, and
$\gamma,\beta$ are complex coefficients 
which include the loop suppression factors $(16\pi^2)^{-n}$ and
counting factors $N$, and so whose magnitudes are roughly ${\cal
O}(10^{-4}-1)$.

We are now in a position to write down the full scalar potential
after SUSY-breaking,
including supergravity-mediated soft masses as well as the tadpole
contributions. Begin by considering the contributions to the scalar
potential involving the $F_S$ auxiliary field:
\beq
V_{F_S}=(\beta\eps\mgrav\mpl F_S+h.c.) -|F_S|^2
- \left(F_S\frac{\partial W}{\partial S}+h.c.\right)
\eeq
where the first term is the contribution of the tadpole.
On integrating out all auxiliary fields, one finds that $F_S$
is shifted from its canonical form by the tadpole contribution:
\beq
F_S^\dagger=-\frac{\partial W}{\partial S} +\beta\eps\mgrav\mpl,
\eeq
while all other $F$-terms (\eg\ $F_\Si$ and $F_{H}$)
are canonical. The $D$-terms associated with the
gauge fields also take their canonical forms.

The full scalar potential
after soft SUSY-breaking can then be written:
\bea
V&=&\sum_i m_i^2|\varphi_i|^2
+|\lambda_\Si S|^2\left(|\Si|^2+|\bar\Si|^2\right) 
+|\lambda_H S|^2 \left(|H|^2+|\bar H|^2\right) \nonumber \\
& &{}+\mgrav^2\mpl(\gamma S+\gamma^\dagger S^\dagger)  
+\left|\lambda_\Si\Si\bar\Si+\lambda_H H\bar H-\beta\eps\mgrav\mpl\right|^2 \\
& &{}+\frac{g_\Si^2}{2}\left(|\Si|^2-|\bar\Si|^2\right)^2 
+\frac{g_H^2}{2}\left(|H|^2-|\bar H|^2\right)^2
+ A,B\mbox{-terms}, \nonumber
\label{eq:V}
\eea
where the first term represents the
gravitationally-induced soft SUSY-breaking masses, $m_i^2\sim\mgrav^2$, for the
fields $\varphi_i=\lbrace S,\Si,\bar\Si,H,\bar H\rbrace$. The final terms are
gravitationally-induced soft-breaking bilinear, $B$, and trilinear, $A$, terms
which for simplicity we ignore hereafter; they do not change our results
substantially. Note that 
the potential as written requires that $m_S^2\geq0$ in order to be bounded 
from below (this condition is modified in the presence of $B$-terms).
Indeed, one expects $m_S^2>0$ at tree level and it will only be
driven negative if 
its coupling to either of the two sets of Higgs fields is fairly large. 
Henceforth we will take all soft squared-masses to be equal to $\mgrav^2>0$.

~

\noindent{\it Minimization of the Potential.}~
To continue further, we take $\eps\simeq1$, \ie, $K_i\simeq\mpl$ which
is the generic choice; small deviations of $\eps$ away from 1 can be
absorbed into $\beta$.
Writing down the minimization conditions for the potential is trivial, but as
the potential is quite complicated, it has many local minima besides the true
global one. However, there are two lowest-lying minima,
both along directions that are $D$-flat up to weak-scale corrections,
\ie, $\Si\simeq\bar\Si$ and $H\simeq\bar H$.

At a first minimum, denoted $V_1$, 
\bea
\Si=\bar\Si=H=\bar H&=&0, \nonumber \\
S&\simeq& -\gamma^\dagger\mpl, \nonumber \\
|F_S|&\simeq&|\beta\mgrav\mpl|, \nonumber \\
V_1\equiv V_{\rm min}&\simeq&(|\beta|^2-|\gamma|^2)\mgrav^2\mpl^2.\nonumber
\eea
This minimum represents the case usually considered in the literature for
singlets with non-zero 
tadpoles --- their vevs are pulled up to the Planck scale, taking with them any
matter to which they couple. This is precisely the reason it was argued in
Refs.~\cite{bp,bpr}\ that the
vev of a true singlet cannot be responsible for the $\mu$-term in the MSSM.

At a second minimum, $V_2$, 
\bea
\Si\bar\Si&=&\frac{\beta\mgrav\mpl}{\lambda_\Si}, \nonumber \\
H=\bar H&=& 0,\nonumber \\
S&\simeq&-\frac{\gamma^\dagger}{2|\lambda_\Si\beta|}\mgrav, \nonumber \\
F_S&\sim&\mgrav^2, \nonumber \\
F_{\Si,\bar\Si}&\simeq&\lambda_\Si\mgrav^{3/2}\mpl^{1/2} ,\nonumber \\
V_2\equiv V_{\rm min}&\simeq&\frac{1}{|\lambda_\Si|}\left(\left|\beta\right|
-\left|\frac{\gamma^2}{2\beta}\right|\right)\mgrav^3\mpl. \nonumber
\eea
The $\Si$-fields receive
vev's of $\sim\sqrt{\mgrav\mpl}$ to cancel off the $F_S$ contribution to the 
potential. These large $\Si$-vev's then produce masses for the $S$-field
(through the $F_\Si$ terms) which stabilizes the $S$-vev against the 
tadpole-induced linear potential. The resulting vev of $S$ is then only
$\langle S\rangle\sim\mgrav\simeq\mw$!

Any gauge symmetry carried by
the $\Si$-fields will be broken at the scale of their vev's. 
Up to the loop factors buried in $\beta$, this is the intermediate scale,
$\mint$. It is also, not
coincidentally, the scale of SUSY-breaking in the hidden sector. In fact, one
may interpret the physics at this minimum as the tadpoles communicating to the
$\Si$-fields the true scale of SUSY-breaking, up to the loop factors.

There is also a third minimum, $V_3$, 
which is identical to $V_2$ except
that the would-be MSSM Higgs fields, $H$ and $\bar H$, 
play the role of $\Si$ and $\bar\Si$ and receive
vev's $\sim\mint$, with $\lambda_H$ replacing $\lambda_\Si$ in all 
expressions. This is clearly {\it not}\/ the desired minimum but 
is instead another 
example of how the tadpole can destabilize the weak scale.
(Note that points at which $H,\bar H,\Si,\bar\Si$ all get vev's
simultaneously are not even local minima of the potential.)

One still needs to resolve which of the three minima is the global one. 
Clearly $|V_{2,3}|\ll|V_1|$, but if
$V_1<0$ then $V_1$ will be the global minimum. It is easy to show that 
$V_1<0$ if and only if $|\gamma|^2>|\beta|^2$. As we know of no general
argument that can set the relative sizes of $\gamma$ and $\beta$, we simply
regard the inequality $|\gamma|^2<|\beta|^2$ to be a condition for
the universe to lie in the interesting minimum, which, given the
arbitrariness of $\beta$ and $\gamma$, is not a restrictive requirement.

But assuming $V_1>V_{2,3}$, which of either pair of fields
$\Si,\bar\Si$ or $H,\bar H$ gets a vev?
Though the form of the potential for the $\Si$- and $H$-fields are the same,
they have differing quadratic pieces (effective masses) coming from their soft
masses and their couplings to $S$. Assuming equality of soft masses, $V$ at its
minimum is positive (given $|\gamma|^2<|\beta|^2$) and scales as either
$|\lambda_\Si|^{-1}$ or $|\lambda_H|^{-1}$. Thus the
global minimum of the potential occurs where only the pair of fields with
largest absolute coupling to $S$ gets vevs $\sim\mint$. That is, if
$|\lambda_H|<|\lambda_\Si|$, it is energetically unfavorable
for $H,\bar H$ to receive vevs $\sim\mint$ and so only the
$\Si,\bar\Si$ fields do. Note in particular that simple inequality is all that 
is needed to ensure that a gauge hierarchy develops; no large 
hierarchy is needed between the two couplings themselves.
Unequal soft masses shift the condition slightly, but the same basic result
will always hold. Thus we conclude that the $S$-vev can in fact provide the
$\mu$-term of the MSSM as long as $|\lambda_H|<|\lambda_\Si|$. 
The whole question of which gauge group is broken at the scale
$\mgrav$ and which at $\mint$ may rest entirely on the relative size of two
couplings ($\lambda_H$ and $\lambda_\Si$) whose ratio is generically 
${\cal O}(1)$!

Recapping, we have found that if $|\beta|^2>|\gamma|^2$ and if $|\lambda_H|<
|\lambda_\Si|$ then $S$ develops a vev at the weak scale and provides a 
$\mu$-term of the correct size for light MSSM Higgs fields. Henceforth we will
assume that these conditions hold and examine the solution in minimum $V_2$.
It is well-known that models
with a dynamical $\mu$-term can contain a
Peccei-Quinn ($PQ$) symmetry which would be spontaneously broken and thus
create an unwanted axion at the weak scale. To examine this possibility,
promote the $PQ$-symmetry to the previously discussed $R$-symmetry under which $S$ is 
neutral and all other superfields are singly charged. 
However, $R(\beta\mgrav\mpl)=2$ explicitly breaks the symmetry and the
would-be axions are all given masses near the intermediate scale, rendering
them harmless. But there still remains a residual $PQ$ symmetry
in the MSSM Lagrangian. This too is explicitly broken, this time by
$F_S\sim\mw^2$, which generates a $B$-term in the Higgs sector 
($\sim F^\dagger_SH\bar H$) and 
gives mass to the pseudoscalar Higgs/would-be axion.

In detail, the heavy spectrum contains two massive pseudoscalars (with 
masses near $\langle\Si\rangle$), a massless Goldstone which is
eaten by the gauge field, and three massive scalars. In particular, the
$S$-scalar itself is not in the light spectrum. It receives a mass
via its coupling to $\Si$ and $\bar\Si$ near $\mint$.
Analysis of the heavy 
fermion mass spectrum is more complicated since $\psi_S$ mixes
with the $\psi_\Si$ and $\psi_{\bar\Si}$ fields and the gaugino,
$\chi$, of the broken gauge group. The mass matrix in the $\{\psi_S,
\psi_\Si, \psi_{\bar\Si}, \chi\}$ basis is given by
\beq
M=\left(\begin{array}{cccc} 0 & \lambda\bar\Si & \lambda\Si & 0 \\
      \lambda\bar\Si & 0 & \lambda S & g\Si \\
      \lambda\Si & \lambda S & 0 & -g\bar\Si \\
      0 & g\Si & -g\bar\Si & m_\chi \end{array}\right)
\eeq
where the (4,4) element is the soft SUSY-breaking
mass of the $\chi$: $m_\chi\simeq\mgrav$. 
This mass matrix has all four eigenvalues of order $\mint$, so
there is no remnant of the dynamical $S$ field at the weak scale, only its weak
scale vev.

While the application of $S$ to the MSSM $\mu$-term does not require
that any of the MSSM fields be charged under the $U(1)_\Si$ gauge group, 
it is easy to see that our mechanism still works even if this charge
were carried by some or all of the MSSM fields. Difficulties could 
arise if the $D_\Si$-term were large after symmetry-breaking,
$D_\Si\gg\mgrav^2$. Then all fields charged under that symmetry would be pulled
up to the scale of the (large) $D_\Si$-term. 
However, the $\Si$-fields receive their
vevs along an approximately $D$-flat direction $\Si\simeq\bar\Si$ where:
\beq
\Si^2-\bar\Si^2=\frac{1}{4g_\Si^2}\left(m_\Si^2-m_{\bar\Si}^2\right).
\eeq
Thus, just as in the case without tadpoles, the mass corrections to the MSSM
fields are not dangerous, though perhaps observable~\cite{dterms}.

It is clear from our analysis why it was necessary to forbid $S^3$ terms
from the superpotential (such a term is automatically forbidden if $S$ carries
zero $R$-charge).
Generically, these terms disrupt minimization of the potential since
$S$ can take on a vev itself to cancel off the large $F_S$ term in $V$. While
this may provide a natural way of giving intermediate scale masses to
vector-like matter, it does not give a $\mu$-term.
(The situation is again somewhat
different in the case of gauge-mediated SUSY breaking~\cite{np}.)

The Lagrangian of Eq.~(\ref{eq:leff}) receives corrections  after SUSY-breaking
from the one-loop effective potential. 
The dominant contribution is an $S^4$ term:
\beq
\Delta
V=\frac{1}{64\pi^2}\mbox{STr}\,M^4\log(M^2/Q^2)\simeq\frac{\lambda_\Si^4}
{64\pi^2}\,\left(\frac{\mgrav}
{\mpl}\right)\,S^4+\cdots .
\eeq
However, this correction is too small to affect the minimization of 
the potential or our results. As well, 
if mixing in $K$ between the hidden and visible
sectors is allowed, terms of the form $K=Z^\dagger Z S/\mpl+h.c.$ 
generically arise. After carrying out the $d^4\theta$ integral, these
echo the form of the tadpole-induced contributions, only with ${\cal O}(1)$
coefficients. Thus their entire effect can be absorbed into redefining
$\beta,\gamma\sim{\cal O}(1)$, which leaves our central result unchanged.

~

\noindent{\it Conclusions.}~
In this letter, we have demonstrated a new mechanism for obtaining 
a weak-scale $\mu$-term in the
MSSM by adding a total singlet in conjunction with a new gauge interaction and
its accompanying Higgs sector. We have used the tadpoles endemic to models
with singlets to drive the breaking of the new symmetry at
the intermediate scale $\sqrt{\mw\mpl}$, to remove all vestiges of 
the singlet from the low-energy theory, and to render all would-be axions
harmless. In this way we have pointed out a natural
way to tie the introduction of intermediate-scale operators, such as those
that might give mass to a right-handed neutrino, to the solution of the
weak-scale $\mu$-problem. The model itself is highly generic and can 
naturally explain
the hierarchy between the weak and intermediate scales in terms of the
ratio of two couplings which is ${\cal O}(1)$. It seems straightforward
to generalize our discussion to more complete models of
physics at intermediate scales~\cite{future}.

\section*{Acknowledgements}
We would like to thank K.S.~Babu, G.R.~Farrar, S.F.~King and H.-P.~Nilles
for conversations. The work of CK is supported 
by Department of Energy contract DE-FG02-90ER40542 and through 
the generosity of Helen and Martin Chooljian. The work of SP is supported
by the Polish Committee for
Scientific Research, grant 2P03B0412 (1997-8)
and by the US-Poland Maria Sklodowska-Curie Fund MEN/DOE-96-264.
The work of NP is
supported by National Science Foundation grant PHY-94-23002.
We would all like to thank
the Aspen Center for Physics, where this work was initiated, for its
hospitality.

\end{document}